\documentclass[10pt,twocolumn, superscriptaddress, amssymb, amsmath, aps, showpacs, prb]{revtex4-1}
\usepackage{graphicx}
\usepackage[hyperindex]{hyperref}

\newcommand{\eff}{\text{eff}}

\newcommand{\iso}{\text{iso}}

\newcommand{\imp}{\text{imp}}
\newcommand{\cacu}{CaCu$_2$(SeO$_3)_2$Cl$_2$}

\begin{document}

\title{CaCu$_2$(SeO$_3)_2$Cl$_2$: spin-$\frac12$ Heisenberg chain compound
with~complex~frustrated~interchain couplings}

\author{Oleg Janson}
\email{janson@cpfs.mpg.de}
\affiliation{Max Planck Institute for Chemical Physics of Solids, 01187 Dresden, Germany}
\author{Alexander A. Tsirlin}
\email{altsirlin@gmail.com}
\affiliation{Max Planck Institute for Chemical Physics of Solids, 01187 Dresden, Germany}
\author{Elena~S.~Osipova}
\affiliation{Department of Chemistry, Moscow State University, 119991 Moscow, Russia}
\author{Peter~S.~Berdonosov}
\email{berdonosov@inorg.chem.msu.ru}
\affiliation{Department of Chemistry, Moscow State University, 119991 Moscow, Russia}
\author{Andrei~V.~Olenev}
\affiliation{SineTheta Ltd., MSU build 1-77, 119991 Moscow, Russia}
\author{Valery~A.~Dolgikh}
\affiliation{Department of Chemistry, Moscow State University, 119991 Moscow, Russia}
\author{Helge Rosner}
\email{rosner@cpfs.mpg.de}
\affiliation{Max Planck Institute for Chemical Physics of Solids, 01187 Dresden, Germany}

\date{\today}

\begin{abstract}
We report the crystal structure, magnetization measurements, and
band-structure calculations for the spin-$\frac12$ quantum magnet
CaCu$_2$(SeO$_3)_2$Cl$_2$.  The magnetic behavior of this compound is
well reproduced by a uniform spin-$\frac12$ chain model with the
nearest-neighbor exchange of about 133\,K. Due to the peculiar crystal
structure, spin chains run in the direction almost perpendicular to the
structural chains. We find an exotic regime of frustrated
interchain couplings owing to two inequivalent exchanges of 10\,K each.
Peculiar superexchange paths grant an opportunity to investigate 
bond-randomness effects under partial Cl--Br substitution.
\end{abstract}

\pacs{75.30.Et, 75.50.Ee, 71.20.Ps, 61.66.Fn}
\maketitle

\section{Introduction}
Low-dimensional magnets remain an attractive playground to study
quantum phenomena\cite{balents} and to understand strongly correlated
electronic systems on a model level.\cite{lee2008} Theoretical
investigations disclose interesting features of numerous simple spin
networks, such as a diamond chain,\cite{takano1996} a kagom\'e
lattice,\cite{hastings2000} or a pyrochlore lattice.\cite{canals1998}
The transfer of these spin lattices to real systems and the subsequent
experimental verification of theoretical results are, however, rather
problematic and stimulate extensive studies of low-dimensional (or
potentially low-dimensional) magnetic materials. Most of these studies
are focused on Cu$^{+2}$ compounds, because the $d^9$ nature and the
pronounced Jahn-Teller effect of Cu$^{+2}$ ion lead to insulating
spin-$\frac12$ compounds with diverse spin-lattice geometries.

Aiming to find hitherto unexplored examples of low-dimensional spin
systems, we investigate copper selenite-chlorides. These compounds
combine two important structural ingredients that lead to unusual physical behavior: i)
SeO$_3$ selenite groups contain the lone-pair Se$^{+4}$ cation that
induces exotic (and potentially polar) crystal structures, as in the
piezoelectric ferrimagnet Cu$_2$OSeO$_3$ (Refs.~\onlinecite{cu2oseo3,
*cu2oseo3_nmr});
ii) Cl atoms show strong hybridization with Cu $3d$ orbitals and mediate
long-range exchange couplings, leading to highly entangled spin lattices
in spin-tetrahedra compounds Cu$_2$Te$_2$O$_5$X$_2$ (X = Cl,
Br)\cite{zaharko2006} showing incommensurate magnetic order, or in the
intricate spin-dimer system (CuCl)LaNb$_2$O$_7$
(Ref.~\onlinecite{cucl_2009, cucl}). 

Here, we present an experimental and computational study of
CaCu$_2$(SeO$_3)_2$Cl$_2$. X-ray diffraction, magnetization measurements, and
band structure calculations are applied to elucidate crystal structure,
electronic structure, and magnetic behavior of this compound. In our study, the complex
structural arrangement of Cu polyhedra is readily disentangled by a microscopic
approach. We find the conventional $x^2-y^2$ orbital ground state for both Cu
sites, and establish a minimum magnetic model of uniform spin-1/2
chains with weak, frustrated interchain couplings.

\begin{figure}[!hb]
\includegraphics[width=8.6cm]{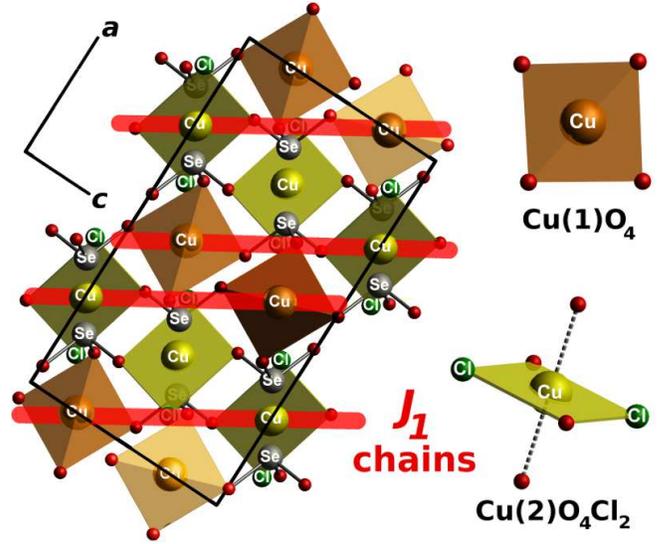}
\caption{\label{F_str}(Color online) Left: crystal structure of \cacu. The structural
chains run along $[10\bar 1]$ (not shown), whereas the magnetic chains
range along $\sim[201]$, as shown by bold (red) lines. Small unlabeled  spheres denote O atoms.
Right: local environment of Cu(1) and Cu(2). The magnetically active Cu(1)O$_4$ and
Cu(2)O$_2$Cl$_2$ plaquettes are highlighted.}

\end{figure}

\section{\label{S-synth}Synthesis and sample characterization}
Calcium selenite CaSeO$_3$ was prepared via solution synthesis, as described in
Ref.~\onlinecite{caseo3}. The solutions of calcium nitrate Ca(NO$_3)_2$
(chemically pure) (6.214\,g) and selenous acid H$_2$SeO$_3$ (98\,\%) (4.886\,g)
in a minimal amount of hot distilled water were mixed. The ammonia 1:5 water
solution was added to fix pH of the solution in the range 7$-$8. The fine white
powder was obtained as a precipitate. The precipitate was then dried at
150\,$^{\circ}$C. According to X-ray powder diffraction (XRPD), the obtained
powder was identified as a hydrate CaSeO$_3\cdot $H$_2$O. The hydrate was
further calcined on a gas burner in a ceramic plate for 30\,min. The resulting
product was identified as a single phase CaSeO$_3$ [space group $P2_1/n$,
$a$\,=\,6.399(5)\,\r A, $b$\,=\,6.782(4)\,\r A, $c$\,=\,6.682(8)\,\r A,
$\beta$\,=\,102.84(6)\,$^{\circ}$]. 

SeO$_2$ was obtained from H$_2$SeO$_3$ by its decomposition under vacuum at
60\,$^{\circ}$C and the sublimation of the resulting substance in a flow of
anhydrous air and NO$_2$. 

CuO (ultra pure) and CuCl$_2$ (Merck, $>$98\,\%) were used. The
dark-greenish powder sample of CaCu$_2$(SeO$_3)_2$Cl$_2$ was obtained
from a stoichiometric mixture of CaSeO$_3$, CuCl$_2$, CuO, and SeO$_2$.
The mixture (about 0.5\,g total) was prepared in Ar-filled camera,
sealed in a quartz tube, and placed into the electronically controlled
furnace. The sample was heated from room temperature to 300~$^{\circ}$C
for 12\,hours, exposed at 300\,$^{\circ}$C for 24\,hours, heated up to
500\,$^{\circ}$C for 12\,hours, and exposed at 500\,$^{\circ}$C
for 96\,hours. 

The resulting samples were single-phase, as confirmed by powder x-ray
diffraction (STOE STADI-P diffractometer, CuK$_{\alpha1}$ radiation,
transmission geometry). The powder pattern was fully indexed in the
monoclinic space group $C2/c$ with lattice parameters
$a$\,=\,12.752(3)\,\r A, $b$\,=\,9.036(2)\,\r A, $c$\,=\,6.970(1)\,\r A,
$\beta$\,=\,91.02(1)\,$^{\circ}$.  \cacu\ is rather stable in air,
although a prolonged exposure of about 3~months led to a partial
decomposition towards crystalline CuSeO$_3$$\cdot$2H$_2$O and possible
amorphous products. 

\begin{table}[htb]
\caption{\label{suppl1}
Data collection and structure refinement parameters for CaCu$_2$(SeO$_3)_2$Cl$_2$. 
}
\begin{ruledtabular}
\begin{tabular}{lr}
  Parameter & Value \\ \hline 
  Temperature (K) & 293(2) \\
  Radiation, $\lambda$ (\r A) & MoK$_{\alpha}$, 0.71069 \\
  Space group   & $C2/c$ (No.~15) \\
  $a$ (\r A)    & 12.759(3)       \\
  $b$ (\r A)    & 9.0450(18)        \\
  $c$ (\r A)    & 6.9770(14)       \\
  $\beta$ ($^{\circ}$) & 91.03(3)        \\
  $V$ (\r A$^3$) & 805.1(3)       \\
  $Z$            & 4              \\
  Calculated density (g/cm$^3$) & 4.059      \\
  Absorption coefficient $\mu$ (mm$^{-1}$) & 15.612      \\
  Crystal size (mm) & $0.30\times 0.07\times 0.06$ \\
  Angle range (deg) & $2.76<\theta<31.96$ \\
  Index ranges & $-18{\leq}h{\leq}18$,\\
  & $0{\leq}k{\leq}13$,\\
  & $-10{\leq}l{\leq}2$ \\
  Reflections: total / independent & 1512/1401 \\
  $R_{\text{int}}$ & 0.0171 \\
  Completeness to $\theta$\,=\,31.96\,$^{\circ}$ & 100.0\,\% \\ 
  Absorption correction & $\psi$-scan corrections \\
  Max./min. transmission & 0.392/0.280 \\
  Parameters refined / restraints & 63/0 \\
  Refinement method & full-matrix \\
  & least-squares on $F^2$ \\
  Goodness of fit on $F^2$ & 1.016 \\
  $R_1$, $wR_2$ ($F_o>4\sigma(F_o)$) & 0.0279, 0.0761 \\
  $R_1$, $wR_2$ (all data) & 0.0430, 0.0796 \\
  Extinction coefficient & 0.00085(18) \\
  Largest diff. peak and hole & $0.914, -1.188$ \\
\end{tabular}
\end{ruledtabular}
\end{table}

\section{\label{S-str}Crystal structure}
For the structure determination, a single crystal was picked up from the
bulk polycrystalline sample. The data were collected at the CAD-4
(Nonius) diffractometer (MoK$_{\alpha}$ radiation) at room temperature.
The analysis of systematic extinctions unambiguously pointed to the
space group $C2/c$ (15).  The lattice parameters $a=12.759(3)$\,\AA,
$b=9.0450(18)$\,\AA, $c=6.9770(14)$\,\AA~and
$\beta=91.03(3)$\,$^{\circ}$ were refined, based on 24 well-centered
reflections in the angular range
$12.01$\,$^{\circ}$\,$<$\,$\theta$\,$<$\,15.84\,$^{\circ}$.  The
diffraction data were collected in an $\omega-2\theta$ mode with the
data collection parameters listed in Table~\ref{suppl1}. A semiempirical
absorption correction was applied to the data based on $\psi$-scans of
seven reflections with $\lambda$ angles close to 90\,$^{\circ}$.

Positions of metal and selenium atoms were found by direct methods
(\textsc{shelxs-97}).\cite{shelx} Oxygen atoms were localized by a
sequence of least-square cycles and difference Fourier syntheses
$\Delta\rho(x,y,z)$. The final refinement with anisotropic atomic
displacement parameters was based on $F^2$ (\textsc{shelxl-97}).\cite{shelx}
Further information and the refinement residuals are given in
Table~\ref{suppl1}. Atomic coordinates and atomic displacement
parameters are listed in Table~\ref{suppl2}.\footnote{Details of the
crystal structure investigation may be obtained from
Fachinformationszentrum Karlsruhe, 76344 Eggenstein-Leopoldshafen,
Germany (ICSD reference number 422179).}

\begin{table}[tb]
\caption{\label{suppl2}
Atomic coordinates and atomic displacement parameters $U_{\iso}$.}
\begin{ruledtabular}
\begin{tabular}{lccccr}
  Atom & Position & $x/a$     & $y/b$     & $z/c$     & $U_{\iso}$
\footnote{The $U_{\iso}$ parameters are given in 10$^{-2}$~\r A$^2$ and obtained as one third of the trace of the orthogonalized $U_{ij}$ tensor.} \\ \hline
  Ca   &  $4e$    & 0         & 0.8425(1) & 0.25      & 1.1(1) \\
  Cu(1)&  $4e$    & 0         & 0.6221(1) & 0.75      & 1.2(1) \\
  Cu(2)&  $4c$    & 0.25      & 0.25      & 0         & 1.2(1) \\
  Se   &  $8f$    & 0.1704(1) & 0.6053(1) & 0.0878(1) & 0.8(1) \\
  Cl   &  $8f$    & 0.6170(1) & 0.5817(1) & 0.1048(1) & 1.7(1) \\
  O1   &  $8f$    & 0.1609(2) & 0.7153(3) & 0.2852(4) & 1.4(1) \\
  O2   &  $8f$    & 0.0493(2) & 0.6598(3) & 0.0070(3) & 1.3(1) \\
  O3   &  $8f$    & 0.1455(2) & 0.4342(3) & 0.1704(4) & 1.4(1) \\
\end{tabular}
\end{ruledtabular}
\end{table}

The crystal structure of CaCu$_2$(SeO$_3)_2$Cl$_2$ is shown in
Fig.~\ref{F_str}. It comprises two inequivalent Cu positions
(Table~\ref{suppl2}).
The Cu(1) atoms show a slightly distorted square-planar Cu(1)O$_4$ environment,
typical for Cu$^{+2}$ oxides (Table~\ref{distances}). In contrast, Cu(2) is
six-fold coordinated, with four oxygens and two chlorines forming an
octahedron, squeezed along Cu(2)--O1 (Table~\ref{distances}). Although such
octahedral coordination of Cu(2) is rather unusual, it still allows for a
square-planar-like crystal-field splitting of $3d$ levels and the conventional
non-degenerate $3d^9$ orbital ground state with the half-filled $3d_{x^2-y^2}$
orbital lying in the plane of a Cu(2)O$_2$Cl$_2$ plaquette
(Fig.~\ref{F_str}).\footnote{In this local coordinate system, $x$ corresponds
to the axis, along which the CuO$_6$ octahedra are squeezed. Consequently, the
$z$ axis runs toward O3 atoms.} This plaquette is formed by two Cu(2)--Cl bonds
and two Cu(2)--O1 bonds. The formation of the plaquette can be qualitatively
understood in terms of different ionic radii for oxygen and chlorine. The
larger size of the Cl atoms makes their effect on the Cu $3d$ orbitals similar
to the effect of O1 with shorter distances to Cu. The resulting crystal-field
splitting resembles that of a CuO$_4$ plaquette and drives one of the atomic
$d$ orbitals half-filled as well as magnetic, that is $a$~$posteriori$
confirmed by our DFT calculations (Sec.~\ref{S-DFT}). 

\begin{table}[tb]
\caption{\label{distances}
Selected interatomic distances (in~\r A) in the CaCu$_2$(SeO$_3)_2$Cl$_2$ structure.
}
\begin{ruledtabular}
\begin{tabular}{llp{.03\textwidth}ll}
Atom pair & Distance & & Atom pair & Distance \\ \hline
Ca--O1 ($\times{}2$)& $2.362(2)$ & &  Cu(1)--O2 ($\times{}2$) & $1.920(2)$ \\
Ca--O2 ($\times{}2$)& $2.458(3)$ & &  Cu(1)--O3 ($\times{}2$)& $2.012(2)$ \\
Ca--Cl ($\times{}2$)& $2.827(1)$ & &         &                \\
Ca--Cl ($\times{}2$)& $2.948(1)$ &  &  Cu(2)--O1 ($\times{}2$)& $1.891(2)$ \\
Se--O1 & 1.705(2)            &   &     Cu(2)--O3 ($\times{}2$)& $2.455(2)$ \\
Se--O2 & 1.707(2)            &   &     Cu(2)--Cl ($\times{}2$)& $2.404(1)$ \\
Se--O3 & 1.684(3)            &   &         &                \\
\end{tabular}
\end{ruledtabular}
\end{table}

In general, the formation of CuO$_2$Cl$_2$ plaquettes is typical for
copper oxychlorides.\cite{cucl,cucl2} However, a unique feature of \cacu\ is the
presence of two longer Cu(2)--O3 bonds which look similar to the
Cu(2)--Cl bonds in terms of interatomic distances but are essentially
inactive with respect to the magnetism, as will be shown in Sec.~\ref{S-magn}.

The Cu(1)O$_4$ plaquettes and the Cu(2)O$_4$Cl$_2$ octahedra share
corners and form chains along $[10\bar{1}]$. However, the bridging O3
atoms do not belong to the Cu(2)O$_2$Cl$_2$ plaquettes, hence a simple
Cu(1)--O3--Cu(2) superexchange is unlikely. Instead, the leading exchange
couplings should run via SeO$_3$ trigonal pyramids which join the
plaquettes into a framework. The Cl atoms shape tunnels that run along
$c$ and accommodate the Ca cations.  Surprisingly,
CaCu$_2$(SeO$_3)_2$Cl$_2$ bears no relation to SrCu$_2$(SeO$_3)_2$Cl$_2$
(Ref.~\onlinecite{srcu}) owing to the smaller size of Ca and the high
flexibility of Cu--Se--O--Cl framework. The arrangement of polyhedra
does not resemble any known structure type either.

\section{\label{S-magn}Magnetic properties}
Magnetic susceptibility ($\chi$) of CaCu$_2$(SeO$_3)_2$Cl$_2$ was
measured with a Quantum Design MPMS SQUID magnetometer in the temperature range $2-380$\,K
in applied fields $\mu_0H$ of 0.5, 2, and 5\,T.

\begin{figure}[tb]
\includegraphics[angle=270,width=8.6cm]{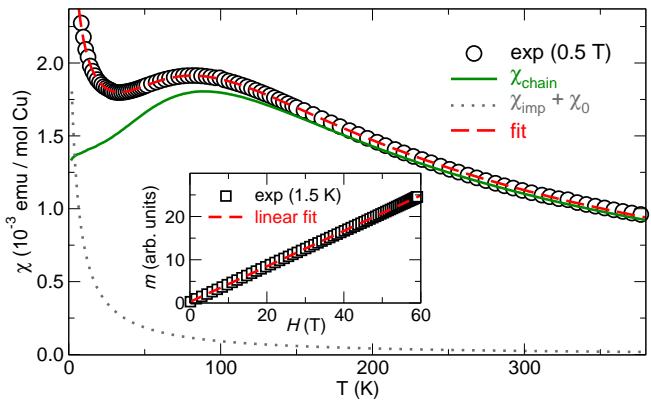}
\caption{\label{F_chiT_MH} (Color online) Magnetic susceptibility (circles) and the fit
(dashed line), see text for details. Intrinsic, i.e. Heisenberg chain
(solid line) and impurity (dotted line)
contributions to the fitted curve are shown. Inset: high-field
magnetization curve (squares) with a linear fit (dashed line).}
\end{figure}

The $\chi(T)$ dependence (Fig.~\ref{F_chiT_MH}) shows a broad maximum at
83\,K and a pronounced increase below 30\,K. The susceptibility maximum
is a signature of quantum fluctuations (low-dimensional and/or
frustrated behavior), while the low-temperature upturn is caused by the
paramagnetic contribution of defects and impurities. Above 230\,K, we
fit the data with the modified Curie-Weiss law
$\chi$\,=\,$\chi_0+C/(T+\theta)$ where
$\chi_0$\,=\,6(1)$\times$10$^{-5}$\,emu\,(mol\,Cu)$^{-1}$ accounts for the
temperature-independent (e.g., van Vleck) contribution,
$C$\,=\,$0.42(1)$\,emu\,K\,(mol\,Cu)$^{-1}$ is the Curie constant, and
$\theta$\,=\,93(5)\,K is the Weiss temperature.
The positive $\theta$ indicates
predominant antiferromagnetic (AFM) interactions in the system. 
Using the expressions
\begin{equation}
C = \frac{N_A\mu_{\text{eff}}^2}{3k_B}, \hskip .05\textwidth \mu_{\text{eff}}= g\mu_B\sqrt{S(S+1)}
\end{equation}
we obtain the resulting effective magnetic moment
$\mu_{\eff}$\,=\,$1.83(1)$\,$\mu_B$ and the $g$-factor $g$\,=\,$2.11(1)$, typical
for spin-$\frac12$ Cu$^{+2}$.\cite{HC_AgCuVO4} 

To fit the whole $\chi(T)$ dependence, we used different expressions for
simple low-dimensional spin models. The best fit was obtained with the
expression for the uniform \mbox{spin-$\frac12$}
chain $\chi_{\text{ch}}$, given by Ref.~\onlinecite{johnston2000} $[$see their Eq.~(53)
parameterized with the values
provided in the third column of Table~I$]$. The temperature range 2\,$-$\,380\,K
fits to the validity condition of this parameterization
0\,$\leq$\,$J$\,$\leq$5.\cite{johnston2000} To account for
temperature-independent and the low-temperature impurity
contribution  to $\chi(T)$, we supplemented $\chi_{\text{ch}}$ with the
$\chi_0$ term and the Curie term $C_{\imp}/T$, respectively:
\begin{equation}
\label{E-fit}
\chi(T) = \chi_0+\dfrac{C_{\imp}}{T}+\frac{N_{A}g^2\mu_{B}^2}{J}\,\chi_{\text{ch}}\left(\frac{T}{J}\right).
\end{equation}
The fit yielded the intrachain exchange coupling $J$\,=\,133(1)\,K, the
$g$-factor $g$\,=\,2.11(1), $\chi_0$\,=\,3(1)$\times$10$^{-5}$\,emu\,(mol\,Cu)$^{-1}$,\footnote{The $\chi_0$ value obtained by fitting
using Eq.~(\ref{E-fit}) is almost
twice smaller than the value obtained from the Curie-Weiss fit. The
reason for this discrepancy is the additional term $C_{\imp}/T$ in
Eq.~\ref{E-fit}. Since (i) this term is of the same order as $\chi_0$
in the high-temperature region, and (ii) both $\chi_0$ and $C_{\imp}/T$ are
positive, $\chi_0$ from the
Curie-Weiss fit is substantially larger than $\chi_0$ from Eq.~\ref{E-fit}.}  and
$C_{\imp}$\,=\,0.005(1)\,emu\,K\,(mol Cu)$^{-1}$ (about 1\,\% of
spin-$\frac12$ impurities). To check the applicability of the Heisenberg chain model to our
system, we calculate
$\chi^{\text{max}}_{\text{ch}}T^{\text{max}}g^{-2}$, that should
amount to 0.0353229(3)\,emu\,K\,(mol\,Cu)$^{-1}$ 
for a Heisenberg chain system, independent of $J$ (see
Eq.~31 from Ref.~\onlinecite{johnston2000}). For \cacu,
$\chi_{\text{ch}}(T^{\text{max}})T^{\text{max}}g^{-2}$\,=\,$0.0345(8)$\,emu\,K\,(mol\,Cu)$^{-1}$
deviates only by few percent from the ideal value, justifying the
mapping onto the Heisenberg spin chain model. 

\begin{figure}[tb]
\includegraphics[width=8.6cm]{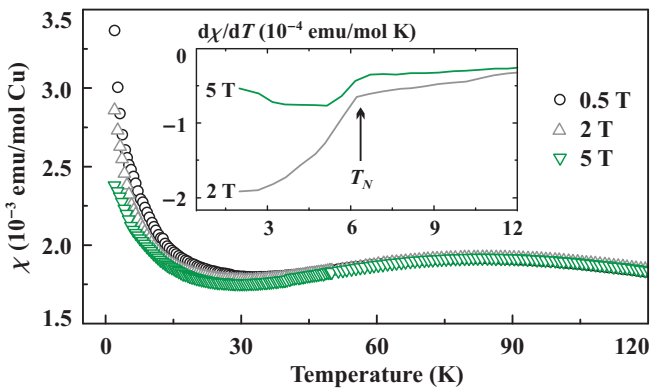}
\caption{\label{F_susc_fields}(Color online) Magnetic susceptibility of
CaCu$_2$(SeO$_3$)$_2$Cl$_2$ measured in applied fields of 0.5\,T, 2\,T, and
5\,T. The increase in the field leads to suppression of the low-temperature
paramagnetic upturn. The inset shows the derivative of the magnetic
susceptibility and a kink around 6\,K, likely evidencing long-range magnetic
ordering.}
\end{figure}

The extrinsic nature of the low-temperature Curie tail is supported by
its suppression in magnetic field (Fig.~\ref{F_susc_fields}). Temperature
derivative of magnetic susceptibility exhibits a kink at 6\,K, which is
likely a signature of antiferromagnetic ordering. This issue is
discussed in context of the microscopic spin model in Sec.~\ref{S-disc}. 

We also measured the magnetization curve of \cacu\ in pulsed magnetic
fields up to 60\,T at a constant temperature of 1.5\,K.\cite{notemh} The
linear change in the magnetization (see the inset of
Fig.~\ref{F_chiT_MH}) is consistent with the proposed uniform-chain
behavior, since the accessible field range is well below the saturation
field ($\mu_0H_s$\,=\,188\,T for $J$\,=\,133\,K and $g$\,=\,2.11) for the parameters
obtained above.\footnote{Our setup did not allow to control precisely
the amount of sample in the coil. Therefore, the magnetization
is measured in arbitrary units. We attempted to scale the high-field curve
using the low-field MPMS data (up to 5\,T), but the resulting
uncertainties impeded obtaining quantitatively consistent results.} 

According to Sec.~\ref{S-str}, the spin chains cannot be assigned to the
structural chains of the corner-sharing Cu(1)O$_4$ plaquettes and
CuO$_4$Cl$_2$ octahedra. In Sec.~\ref{S-DFT}, we show that the uniform
chains originate from a nontrivial superexchange pathway via SeO$_3$
groups.

\section{\label{S-DFT}Microscopic model}
Band structure calculations have been performed using the full-potential
code \textsc{fplo9.00-31}.\cite{fplo} For the exchange and correlation potential,
the parameterization of Perdew and Wang has been chosen.\cite{pw} For the
calculations within the local density approximation (LDA), a well
converged $k$-mesh of 10$\times$10$\times$12 points was used. For
spin-polarized supercell local spin density approximation (LSDA)+$U$ calculations, we used $k$-meshes of
4$\times$4$\times$4 and 4$\times$4$\times$2 points. Convergence of total
energy with respect to the $k$-mesh has been carefully checked.

\begin{figure}[tb]
\includegraphics[angle=270,width=8.6cm]{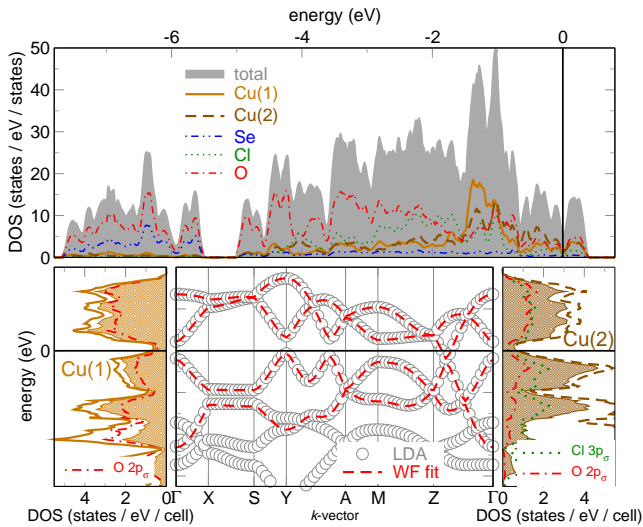}
\caption{\label{F_dosband} (Color online) Top: total (shaded) and atomic-resolved
(lines) density of states (DOS) for \cacu. Bottom center: LDA band structure
(circles) and the WF fit (dashed line). Bottom left and right: orbital-resolved
density of states for Cu(1) and Cu(2).  Solid $[$dashed$]$ lines denote total
DOS for Cu(1) $[$Cu(2)$]$. Shaded regions show the $3d_{x^2-y^2}$ contribution.
O 2p$_{\sigma}$ (Cl 3p$_{\sigma}$) DOS are shown with dash-dotted (dotted)
lines.
}
\end{figure}

LDA yields a valence band (Fig.~\ref{F_dosband}, top) with the bandwidth
of about 8\,eV, typical for cuprates.\cite{bicu, cuse, cu2v2o7, volb} The
band is clearly split into two parts: the region between $-8$ and
$-5.5$\,eV is dominated by Se $4p$ and O $2p$ states, while the rest of the valence
band is formed by Cu, O, and Cl states. Non-zero density of states (DOS)
at the Fermi level $\varepsilon_{\text{F}}$ indicates a metallic
solution in contrast to the insulating behavior, expected from the
green sample color. This well-known drawback of the LDA arises from a strong
underestimation of correlations which are intrinsic for the $3d^9$
electronic configuration (Cu$^{2+}$) and drive the system into the
insulating regime.\cite{cupr}

The states relevant for magnetism are confined to the vicinity of
$\varepsilon_{\text{F}}$. In most cuprates, these are the antibonding Cu
$3d_{x^2-y^2}$ and O $2p_{\sigma}$ states (in the local coordinate
system), typically well-separated from the rest of the valence
band.\cite{cuorg, bicu, cuse, cucl}  However, \cacu\ lacks a
separated band complex around $\varepsilon_{\text{F}}$. This reflects
the octahedral coordination for Cu(2) and makes a detailed analysis of the
magnetically active orbitals necessary. 

\begin{figure}[tb]
\includegraphics[width=8.6cm]{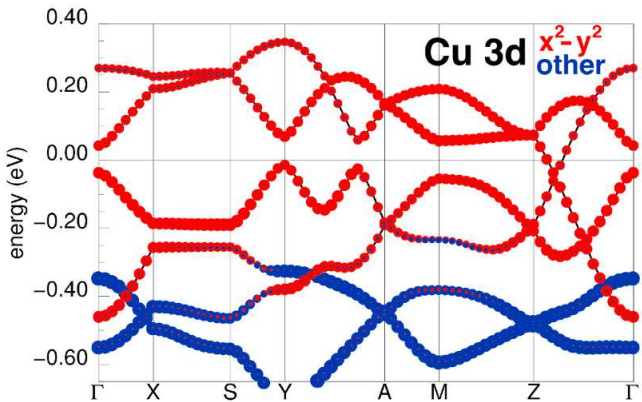}
\caption{\label{F_bandw}(Color online) Band weights for the magnetically active
Cu $3d_{x^2-y^2}$ and other Cu $3d$ orbitals.}
\end{figure}

To evaluate the relevant states, we project the DOS onto a set of
local orbitals. This way, we find the dominant $3d_{x^2-y^2}$
contribution to the Cu(1) DOS at $\varepsilon_{\text{F}}$
(Fig.~\ref{F_dosband}, left bottom), as for almost all undoped cuprates. For the Cu(2)
atom, the situation is less trivial, since the local environment of this
atom implies two short Cu(2)--O bonds as well as four long (two Cu(2)--O
and two Cu(2)--Cl) almost equidistant bonds, making the choice of the
local coordinate system ambiguous. However, the analysis of local DOS for
different situations readily yields the correct choice of the local axes
and evidences that the two short Cu(2)--O and two Cu(2)--Cl bonds form a
plaquette with the Cu $3d_{x^2-y^2}$ magnetically active orbital
(Fig.~\ref{F_bandw}). The local DOS of this orbital clearly
dominates the states at $\varepsilon_{\text{F}}$  (Fig.~\ref{F_dosband},
right bottom) and confirms our empirical considerations presented in
Sec.~\ref{S-str}.

Since the magnetism of \cacu\ is confined to Cu $3d_{x^2-y^2}$
orbitals, they can be used as a minimal basis for an effective
tight-binding (TB) model.  The total number of states (four) in the
model corresponds to the number of magnetic Cu atoms in a unit cell: two
Cu(1) and two Cu(2). To parameterize the model, we use the Wannier
functions (WFs) technique, which yields numerical values for the leading
hoppings $t_i$ (transfer integrals). This way, we obtain a perfect fit
to the LDA bands (Fig.~\ref{F_dosband}, bottom).

\begin{figure}[tbp]
\includegraphics[width=8.6cm]{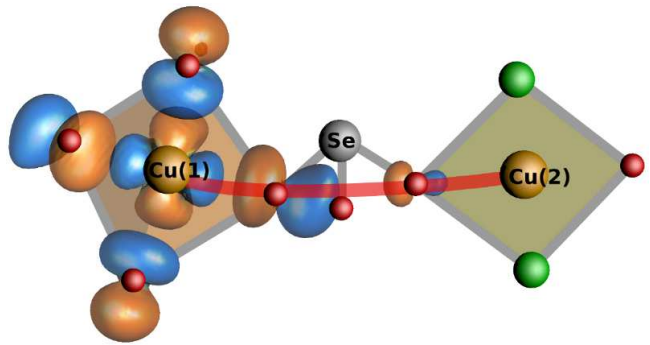}
\caption{\label{F_wf}(Color online) Wannier functions marking the $J_1$ superexchange
path. Cu(1)O$_4$ (left) and Cu(2)O$_2$Cl$_2$ (right) plaquettes are filled. The SeO$_3$
pyramid is visualized by Se--O bonds (lines). The projection is the same
as in Fig.~\ref{F_str} (left).}
\end{figure}

Within the effective one-orbital TB model, we find three relevant
couplings (Table~\ref{dtJ}): $t_1$ (Figs.~\ref{F_wf} and \ref{F_model})
running along Cu(1)-Cu(2) chains almost parallel to the $[201]$
direction, the short-range interchain coupling $t_{\text{ic}1}$ which
connects Cu(1) atoms, as well as the long-range interchain coupling
$t_{\text{ic}2}$ connecting Cu(2) atoms along $[10\bar{1}]$. The
clearly dominant $t_1$ amounts to 139\,meV, while $t_{\text{ic}1}$ and
$t_{\text{ic}2}$ are found to be 47 and 30\,meV, respectively. Due to
the particular orientation of magnetically active orbitals, the
hopping $t_{\text{nn}}$ along the ``structural chains'' is apparently
small, does not affect the magnetic ground state, as will be shown
later.

\begin{table}[tb]
\caption{\label{dtJ}
Interatomic Cu--Cu distances ($d$, in~\r A), transfer integrals
($t_i$, in~meV), as well as antiferromagnetic (AFM) and ferromagnetic (FM)
contributions to the total exchange integrals ($J_i$, in~K) for the leading
couplings in \cacu. For notation of the paths see Fig.~\ref{F_model}
($J_{\text{nn}}$ stands for the nearest-neighbor coupling along the
structural chains).}
\begin{ruledtabular}
\begin{tabular}{c c c c c c c}
path & atoms & $d$ & $t_i$ & $J^{\text{AFM}}_i$ & $J^{\text{FM}}_i$ & $J_i$ \\ \hline
$J_{\text{nn}}$   & Cu(1), Cu(2) & 3.84 &  19 &   4 &   $-$ & 4   \\   
$J_{\text{ic}1}$ & Cu(1), Cu(1) & 4.13 &  47 &  25 & $-15$ & 10  \\   
$J_1$             & Cu(1), Cu(2) & 6.19 & 139 & 200 & $-55$ & 145 \\   
$J_{\text{ic}2}$ & Cu(2), Cu(2) & 7.33 &  30 &  10 &   $-$ & 10  \\   
\end{tabular}
\end{ruledtabular}
\end{table}

To restore the insulating ground state, we map our TB model onto a
Hubbard model with an effective on-site Coulomb repulsion $U_{\eff}$.
Since the strongly correlated limit ($U_{\text{eff}}$\,${\gg}$\,$t_i$)
and the half-filling are well-justified for undoped cuprates, the
low-energy (magnetic) excitations of the Hubbard model can be well
described by a Heisenberg model.  Within the second-order perturbation
theory, AFM exchange couplings are expressed as
$J^{\text{AFM}}_i$\,=\,$4t^2_i/U_{\text{eff}}$.  \footnote{The
second-order perturbation theory accounts only for the effective
hoppings between the source and target magnetic atoms. Therefore,
$t_{ij}$ is a resulting hopping that contains many individual hopping
process, including hopping to and between the ligand orbitals.} Adopting
a typical value $U_{\text{eff}}$\,=\,4.5\,eV,\cite{cuse,bicu,cucl} we
obtain $J^{\text{AFM}}_1$\,=\,200\,K,
$J^{\text{AFM}}_{\text{ic}1}$\,=\,25\,K and
$J^{\text{AFM}}_{\text{ic}2}$\,=\,10\,K.

The one-orbital approach yields only the AFM part $J^{\text{AFM}}_i$ of
the total exchange $J_i$. However, particular geometrical configurations
(Cu--O--Cu angle close to 90$^{\circ}$, edge-sharing of CuO$_4$
plaquettes etc.) often lead to sizable ferromagnetic (FM) contributions to the
total exchange, which can even outgrow $J^{\text{AFM}}_i$,
resulting in a FM (negative) $J_i$. Based on structural considerations
only, an appreciable FM contribution might
be expected for the short-range coupling $J_{\text{ic}1}$ and the
nearest-neighbor coupling $J_{\text{nn}}$, whereas
$J_1$ and $J_{\text{ic}2}$ are rather long-range, and their FM
contributions should be small. To challenge this conjecture, we perform
supercell LSDA+$U$ calculations. This method gives access to the total
exchange $J_i$, being the sum of the AFM and FM contributions,
$J^{\text{AFM}}_i$ and $J^{\text{FM}}_i$, respectively. Combining the
LSDA+$U$ results with $J^{\text{AFM}}_i$ estimates from the model
approach described above, the FM contributions $J^{\text{FM}}_i$ can be
evaluated.

Recent thorough studies on low-dimensional cuprates give evidence that
quantitative magnetic models based on the results of LSDA+$U$
calculations are, in addition to the dependence on the Coulomb repulsion
$U_d$, also dependent on the double-counting correction (DCC)
scheme.\cite{cu2v2o7,volb} The influence of these two parameters is
particularly large for systems with sizable $J^{\text{FM}}_i$. Yet, for
cuprates with structurally isolated plaquettes, as in the case of \cacu, the
around-mean-field (AMF) DCC with the $U_d$ value 6.5\,$\pm$\,0.5\,eV
typically yields accurate ($\pm10\%$) estimates for the leading
couplings.  Adopting the AMF DCC and $U_d$\,=\,6.5\,eV, we obtain
$J_1$\,=\,145\,K, in excellent agreement with the experimental
$J$\,$\simeq$\,133\,K from the fit to the magnetic susceptibility. In
accordance with our expectations, the long-range interchain coupling
$J_{\text{ic}2}$ has a tiny FM contribution only, while for the short-range
coupling $J_{\text{ic}1}$ the FM contribution reaches
$J^{\text{FM}}_{\text{ic}1}$\,=\,$-$15\,K (see Table~\ref{dtJ}). At first
glance, the rather large $J^{\text{FM}}_1\sim-55$\,K may look surprising.
However, in the related quasi-1D system CuSe$_2$O$_5$, a similar
intrachain coupling runs via two
corner-sharing SeO$_3$ pyramids and the FM contribution
$J^{\text{FM}}_1$ to this coupling amounts to even $-$100\,K.\cite{cuse}
Therefore, such high FM contributions are likely intrinsic to the
superexchange realized via SeO$_3$ pyramids.  Understanding the
underlying mechanism of this complex superexchange deserves a separate
study and lies beyond the scope of the present investigation. 
Presently, we note that the superexchange results from the overlap of
oxygen orbitals, while Se states have only a minor contribution to the WFs.

The last remark concerns the short-range coupling between the
nearest-neighbor Cu(1) and Cu(2) atoms that form the structural chains.
The WFs analysis yielded a negligible $t$ associated with this coupling
path.  Still, the respective interatomic distance (3.84~\AA) 
is relatively small, which could give rise to a FM coupling.
Therefore, we evaluated the total exchange $J_{\text{nn}}$ using the LSDA+$U$
calculations.  The resulting exchange of 4\,K is in excellent agreement
with the TB estimate (4\,K), evidencing a negligible FM contribution and
justifying our restriction to the three couplings $J_1$,
$J_{\text{ic}1}$, and $J_{\text{ic}2}$ for a minimum model.

\section{\label{S-disc}Discussion and summary}

\begin{figure}[tbp]
\includegraphics[width=8.6cm]{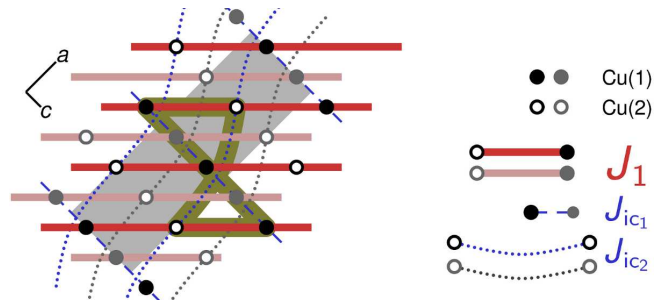}
\caption{\label{F_model} (Color online) Spin model for \cacu. Filled and
empty circles show the Cu(1) and Cu(2) positions. Bold lines and circles
denote couplings (see legend for notation; the same notation applies for Table~\ref{dtJ}) in the
front plane, whereas gray circles and shaded lines correspond to the atoms
lying in the back plane. The planes are connected 
by $J_{\text{ic}1}$ couplings only. A closed loop (bold line)
having an odd number of AFM couplings indicates that the
spin model is frustrated. The unit cell is depicted by the gray rectangle.}
\end{figure}

The spin model of \cacu\ is depicted in Fig.~\ref{F_model}. Its main
element are Cu(1)--Cu(2) chains running almost parallel to the $[201]$
direction, which is different from the structural chains.  The spin-$\frac12$
chains are coupled by two inequivalent exchange interactions:
$J_{\text{ic}1}$ is short-range and links the Cu(1) atoms
of the two neighboring chains, while the long-range $J_{\text{ic}2}$
bridges the Cu(2) atoms of the fourth-neighbor chains. Another
difference between $J_{\text{ic}1}$ and $J_{\text{ic}2}$ is that the
former is responsible for a 3D coupling $[$connects Cu(1) atoms belonging
to different layers, see Fig.~\ref{F_model}$]$, whereas the latter is
confined to the $ac$ plane. Either of inter-chain couplings alone,
$J_{\text{ic}1}$ or $J_{\text{ic}2}$, leads to a
3D or 2D non-frustrated model, respectively.
However, the combination of the two inter-chain couplings gives rise to
magnetic frustration, evidenced by an odd number of AFM bonds along the
closed, hourglass-shaped loop shown in Fig.~\ref{F_model}.

In general, most of the quasi-one-dimensional cuprates order AFM at low temperatures,
while strong quantum fluctuations drastically reduce the value of the
ordered moment compared to the classical value of 1\,$\mu_{\text{B}}$.\cite{*[{See, e.g. }] [{ and references therein.}]
sr2cuo3} The AFM transition temperature can vary in a wide range, since
it depends not only on the leading intrachain coupling $J_1$, but also on
the absolute values and the topology of the interchain couplings.  To
emphasize the huge impact of the interchain coupling regime onto the
transition temperature, we mention here two quasi-one-dimensional
systems with similar intrachain coupling $J_1$ of 150--200\,K,
but different interchain coupling regimes: a frustrated interchain
coupling in Sr$_2$Cu(PO$_4$)$_2$ leads to a very low ordering
temperature of 85\,mK
($T_{N}/J$\,=\,4.5$\times$10$^{-4}$) (Ref.~\onlinecite{sr2cupo42_exp}) despite an
interchain coupling of 3\,K.\cite{sr2cupo42_dft} 
On the contrary, a sizable and non-frustrated interchain coupling
$J_{\text{ic}}$\,$\sim$\,20\,K in CuSe$_2$O$_5$ results in a long-range
AFM ordering at rather high $T_{N}$\,=\,17\,K
($T_{N}/J$\,=\,0.1).\cite{cuse}

Since \cacu\ exhibits a similar energy scale ($J_1$\,$\sim$\,133\,K), it is
natural to compare this compound to the aforementioned systems. The
major difference here is the presence of two types of interchain
couplings, $J_{\text{ic}1}$ and $J_{\text{ic}2}$, which form a 3D spin
model, in contrast to Sr$_2$Cu(PO$_4$)$_2$ and CuSe$_2$O$_5$, where the
leading interchain couplings are confined to 2D, while the coupling
along the third direction $J_{\perp}$ is substantially smaller.  This
argument favors higher $T_N$ in \cacu. On the other hand, the interchain
couplings in \cacu\ are frustrated, which certainly inhibits the magnetic
ordering and can considerably lower the $T_N$. We expect the combination
of the 3D coupling regime and the frustration to result in a moderate
$T_N$ of \cacu, comparable to that of CuSe$_2$O$_5$. Indeed, the kink of
magnetic susceptibility at 6\,K in Fig.~\ref{F_susc_fields} fits well
to the energy scale of the anticipated long-range ordering in \cacu\
(compare to $T_N$\,=\,17\,K in CuSe$_2$O$_5$ with non-frustrated
interchain couplings).

In general, magnetic specific heat $C_p^{\text{magn}}$ data provide an
independent estimate for the leading exchange coupling and are sensitive
to the long-range magnetic ordering. However, the measured specific heat
contains, apart from $C_p^{\text{magn}}$, also a phonon contribution.
The energy scale of magnetic interactions in \cacu\ gives rise to a
maximum in $C_p^{\text{magn}}(T)$ at rather high
$T$\,$\approx$\,(0.48\,$J$)\,=\,64\,K (Eq.~39 in
Ref.~\onlinecite{johnston2000}). At this temperature, the phonon
specific heat strongly dominates over $C_p^{\text{magn}}$, impeding
an accurate disentanglement of magnetic and phonon contributions. 
Moreover, the large value of $J$ leads to only a small
amount of magnetic entropy, which could be released at the transition
temperature. Thus, the resulting magnetic specific heat anomaly would be rather
small.\cite{*[{See, e.g. }] [{}] HC_AgCuVO4} Therefore, considering
large quantum fluctuations that substantially lower the ordered magnetic
moment,\cite{sr2cuo3} the method of choice are muon
spin resonance ($\mu$SR) experiments that should be carried out in
future to verify magnetic ordering in \cacu. For instance, long-range
magnetic ordering in the square-lattice compounds
Cu(Pz)$_{2}$(ClO$_{4}$)$_{2}$ and $[$Cu(Pz)$_2$(HF$_2$)$]$BF$_4$ was
only revealed by $\mu$SR, while the heat capacity data lack any
signatures of transition
anomalies.\cite{FSL_CuPz2ClO42_CuPz2HF2BF2_CpT_muSR}

Magnetic frustration is one of the leading mechanisms that give rise to
complex magnetic structures. It is therefore interesting to address the
nature of the anticipated magnetically ordered ground state of \cacu. 
First, we consider the low-energy sector of a classical Heisenberg model
on finite lattices of 16 coupled chains. For each chain, we
impose a condition of the ideal antiferromagnetic arrangement of
the neighboring spins.\footnote{This condition makes the chains
effectively infinite, since the number of possible states for each chain
amounts to two: a certain spin can be up or down, which governs the
arrangement of all other spins in the chain, independent of the chain length.}
To keep the problem computationally feasible, we first restrict
ourselves to collinear
spin arrangements. The magnetic ground state is evalauted as a state with minimal
energy. Adopting the ratios of the leading exchange couplings
from our LSDA+$U$ calculations (Table~\ref{dtJ}, last column), we arrive
at an AFM ground state, with the magnetic unit cell doubled along $a$
and quadrupled along $c$ with respect to the crystallographic unit cell,
i.e. the propagation vector is $(\frac12,0,\frac14)$.  To understand the
particular way the frustration is lifted, we analyse which couplings are
satisfied, by considering the products
$-$4$[\vec{S}_i\cdot\vec{S}_j](J_{ij}/|J_{ij}|)$ for all $(i,j)$ spin
pairs in a unit cell. For collinear
configurations, such product amounts either to 1 (a satisfied coupling)
or to $-1$ (an unsatisfied coupling). For a certain type of exchange
coupling, the sum of such products can be divided by the total number of
couplings of this type in the unit cell (note the multiplicities
different from 1).
This way, we can estimate the fraction of satisfied couplings.  Such
analysis yields that 100\,\% of $J_1$ and $J_{\text{ic2}}$, but only 75\,\% of
$J_{\text{ic1}}$ couplings are satisfied in the proposed $(\frac12,0,\frac14)$
ground state.

Taking into account the restriction to collinear states, it is worth to
address the stability of this ground state using alternative techniques.
Thus, we use a classical Monte-Carlo code\footnote{The simulations are
done for a finite lattice of 48$\times$48$\times$24 spins with periodic
boundary conditions. We use 20000 sweeps for thermalization and 200000
sweeps after thermalization.} from the \textsc{alps} simulations
package\cite{ALPS} and calculate diagonal spin correlations
${\langle}S^z_iS^z_j\rangle$, where $i$ and $j$ are spins coupled by a
particular magnetic exchange.  This way, we obtain $-$0.24452(1),
$-$0.16716(1) and $-$0.23056(3) for the $J_1$, $J_{\text{ic}1}$ and
$J_{\text{ic}2}$ couplings, respectively. These numbers should be
compared to ${\langle}S^z_iS^z_j\rangle$\,=\,$-$0.25 for a perfect
antiferromagnetic arrangement. Despite the small deviations from this
ideal number, the spins coupled by $J_1$ and $J_{\text{ic2}}$ can be
regarded as antiferromagnetically arranged, corroborating our classical
energy minimization result. On the contrary, the value for
$J_{\text{ic1}}$ is substantially smaller, yielding the average angle of
$\arccos{(0.16716/S^2)}$\,$\approx$\,48\,$^{\circ}$ between the respective
spins. The resulting angle is very close to $\pi/4$, hence spins in the
fourth-neighbor chains are almost antiparallel to each other (the angle
amounts to $\pi$). This is in accord with the almost antiparallel
arrangement of spins coupled by $J_{\text{ic2}}$ (coupling
between the fourth-neighbor chains).

In the classical model, the exotic regime of frustrated interchain
couplings leads to a rather complex magnetic ordering in \cacu: the
classical energy minimization yields the collinear $(\frac12,0,\frac14)$
state, while the classical Monte Carlo simulations are in favor of a
non-collinear magnetic ground state. These two ground states differ only
by the mutual arrangement of spins coupled by $J_{\text{ic1}}$.

Since for quasi-one-dimensional systems quantum fluctuations are
crucial, the respective quantum model should be addressed.  However, the
study of a magnetic ordering for a three-dimensional frustrated quantum
magnet is a challenging task, since standard methods, such as exact
diagonalization, quantum Monte-Carlo, and the density matrix
renormalization group technique, are either not applicable or do not
account for the thermodynamic limit. Moreover, \cacu\ features a
non-negligible magnetic impurity contribution, as evidenced by the
low-temperature upturn in $\chi(T)$ (Fig.~\ref{F_chiT_MH}). At low
temperatures, these impurities can give rise to strong internal
fields,\cite{[{Similar effect was recently discussed for the anisotropic
triangular lattice model in the one-dimensional limit, see
}][{}]HC_Cs2CuCl4_theory} and possibly affect the ground state.
Therefore, the magnetic ordering in \cacu\ deserves additional
investigation using alternative techniques, both from the experimental as
well as the theoretical side.

In contrast to well-known uniform-chain systems, \cacu\ shows a complex
crystal structure with two Cu positions revealing an apparently
different local environment. Our DFT calculations evaluate the
$3d_{x^2-y^2}$ magnetically active orbitals lying within the Cu(1)O$_4$ and
Cu(2)O$_2$Cl$_2$ plaquettes. Although the Cu(2)--O3 bond lengths are
similar to those of Cu(2)--Cl bonds, the symmetry of the magnetic
orbitals renders the nearest-neighbor superexchange $J_{\text{nn}}$ path (along
Cu(2)--O3 bonds) essentially inactive (Table~\ref{dtJ}). The spin chains run in a
different direction which is dictated by the orbital state of Cu and by
\cite{}the suitable overlap of the ligand orbitals in the SeO$_3$ groups.  This
nontrivial situation is very typical for spin-1/2 systems
where spin lattices are essentially decoupled from the low-dimensional
features of the crystal structure: recall, for example,
(CuCl)LaNb$_2$O$_7$,\cite{cucl} BiCu$_2$PO$_6$,\cite{bicu2po6}
Cu$_2$(PO$_3)_2$CH$_2$,\cite{cu2po32ch2} and
CuTe$_2$O$_5$.\cite{cute2o5}

Although the intricate regime of the interchain couplings in \cacu\
complicates theoretical studies, the nontrivial structural organization
of this compound also has an important advantage. The Cl and Br atoms
are known to be easily substitutable owing to their similar chemical
nature. The substitution is commonly used to create bond randomness and to access
the exotic behavior of partially disordered spin systems.\cite{[{For
example: }][{}]manaka2008} If the Cl atoms take part in the
superexchange, the chemical substitution inevitably changes the geometry
of the superexchange pathways, and immediately leads to dramatic changes
in the spin system. In \cacu, the Cl atoms lie away from the leading
superexchange pathway (Fig.~\ref{F_wf}), and a moderate alteration of the
magnetism should be expected. Partial Cl/Br substitution will basically modify
the relevant microscopic parameters (such as the crystal-field splitting)
without changing the superexchange geometry.

In summary, we have investigated the crystal structure, electronic
structure, and magnetic behavior of \cacu. The compound comprises two Cu
sites with essentially different local environment, but the same
magnetically active orbital of local $x^2$\,$-$\,$y^2$ symmetry. A peculiar
arrangement of magnetic plaquettes makes \cacu\ a good realization of
the \mbox{spin-$\frac12$} antiferromagnetic Heisenberg chain model with an
intrachain exchange coupling of $\sim$133\,K and frustrated interchain
couplings realized via two inequivalent superexchange paths. A kink in the
magnetic susceptibility at 6\,K hints at long-range magnetic ordering,
which is subject to future experimental verification. The good potential for a
partial substitution of Cl by Br atoms allows to look at the material
from a different point of view. In particular, Cl atoms located close to
but not directly on the leading superexchange path make \cacu\ a
promising model system to study the bond-randomness effects --- like glass
formation --- in low-dimensional magnets.

\acknowledgments
A. T. was funded by Alexander von Humboldt Foundation.  E. O., P. B., A.
O., and V. D. acknowledge the financial support by RFBR in the project
09-03-00799-a. Part of this work has been supported by EuroMagNET II
under the EC contract 228043. We are grateful to Yurii Skourski for his
kind help during the high-field magnetization measurement, and Deepa
Kasinathan for fruitful discussions. We are grateful to Juri Grin for valuable
comments on the manuscript. 

%\bibliography{CaCu2}

\begin{thebibliography}{42}
\bibitem [{not()}]{notemh}%
  \BibitemOpen
  \href@noop {} {}\bibinfo {note} {Detailed description of the measurement
  technique can be found in Ref.~\onlinecite{cu2po32ch2}.}\BibitemShut {Stop}%
\end{thebibliography}
%

\end{document}